\documentstyle[aaspp4]{article} 
\begin{document} 
 
\title{Newly Synthesized Elements and Pristine Dust\\
in the Cassiopeia A Supernova Remnant\footnote{Based on observations
with {\it ISO}, an ESA project with instruments funded by ESA Member States
(especially the PI countries: France, Germany, the Netherlands and
the United Kingdom) and with the participation of ISAS and NASA}} 
\author{R. G. Arendt\altaffilmark{2}, E. Dwek, and S. H. Moseley} 
\affil{Laboratory for Astronomy and Solar Physics,\nl
Code 685, NASA GSFC, Greenbelt, MD 20771;\nl 
arendt@stars.gsfc.nasa.gov, edwek@stars.gsfc.nasa.gov, 
moseley@stars.gsfc.nasa.gov} 
\altaffiltext{2}{Raytheon STX} 
 
\begin{abstract} 
Spectroscopic observations at 2.4 -- 45 $\micron$ of the 
young supernova remnant Cas A with the {\it Infrared Space Observatory} 
({\it ISO}) Short Wavelength Spectrometer (SWS) reveal strong emission 
lines of O, Ne, Si, S, and Ar. These lines are observed at high 
velocities (several 10$^3$ km s$^{-1}$), and are therefore associated with the 
supernova ejecta known as the fast-moving knots (FMKs). Continuum emission from 
dust is also seen in the Cas A spectrum. The continuum strength is spatially 
well correlated with the O and Ar line strengths, indicating that the 
dust emission also arises from the FMKs. The dust continuum has an emission 
feature at $\sim$22~$\micron$ which cannot be fit by typical astronomical 
silicates, but can be fit with a particular class of silicate minerals. 
This suggests that the dust in Cas A is silicate material that has freshly 
condensed from the Cas A ejecta into a mineral form that is uncharacteristic 
of typical ISM dust grains. 
 
\end{abstract} 
 
 
\section{INTRODUCTION} 
Infrared (IR) observations of supernova remnants (SNRs) primarily reveal the 
thermal continuum emission of shock-heated dust (Dwek \& Arendt 1992; and 
references therein). For most remnants, this 
dust is typical interstellar dust that has been swept up in the expanding 
supernova blast wave. Only the very youngest SNRs offer the possibility of 
observing dust that has formed from the metal-enriched ejecta
of the supernova itself, before it is dispersed and mixed into the general ISM. 
The dust content of SNRs is not directly observable in any other wavelength 
regime. The IR portion of the spectrum also contains ground-state 
fine-structure lines from neutral to moderately ionized species of atoms
from carbon to nickel. Observations of these fine structure lines offer 
the capability of probing density and temperature regimes not represented 
by the commonly observed optical transitions. Another distinct advantage 
of IR observations is that extinction is much lower at IR than optical 
or UV wavelengths.

The earliest airborne and ground based searches for IR emission from the Cas A 
supernova remnant (SNR) only managed to set upper limits on the intensity 
of the IR emission (Wright et al. 1980, Dinerstein et al. 1982). 
Observations with the {\it IRAS} satellite, launched in 1983, provided the 
first clear detection of Cas A at IR wavelengths. 
These measurements of the mid-IR emission in 
broad bands at 12, 25, 60 and 100 $\micron$ enabled several detailed 
investigations into the nature of the heating mechanism of the dust within 
the remnant (Dwek et al. 1987a, Braun 1987). 
However the {\it IRAS} data lacked the spatial resolution to 
reveal more than the gross distribution of dust within the SNR. The {\it IRAS} 
data also lacked the spectral resolution needed to determine the amount of 
line emission that might be contributing in the mid-IR. Subsequent observations 
by (Dinerstein et al. 1987) detected weak line emission 
of [S IV] at 10.4 $\micron$. Greidanus \& Strom (1991) 
made a higher resolution ground-based map of the northern part of the 
remnant at 20 $\micron$. This map reveals clumpy IR emission that does 
not appear to be strongly correlated with either the optical or X-ray emission. 
With the advent of the {\it ISO}, IR observations with high spatial 
and spectral resolution are available. An ISOCAM map of Cas A at 
10.7 -- 12 $\micron$ (Lagage et al. 1996) shows emission 
that is distributed in a manner similar to the de-extincted X-ray emission, or 
the radio emission. Much of the brightest emission correlates with line 
emission from high velocity ejecta within the SNR, leading Lagage et al. 
(1996) to propose that the observed emission arises from dust that has
formed in the knots of supernova ejecta, and is now being heated as 
the knots are evaporating into the hot gas of the supernova blast wave 
(Dwek \& Werner 1981).

In this paper we present analysis of IR observations of the young Cas A 
SNR. The 2.4 -- 45 $\micron$ data acquired with the {\it ISO} Short Wavelength 
Spectrometer are described in \S 2. Section 3 is devoted to analysis
of the observed line and continuum emission. Section 4 discusses the 
implications for mixing of the supernova ejecta, and the formation of dust 
within the ejecta.
 
\section{DATA AND REDUCTION} 
\begin{deluxetable}{ccccc}
\tablecaption{Observed Locations\label{Table 1}} 
\tablewidth{0pt} 
\tablehead{ 
\colhead{Field}& 
\colhead{$\alpha$ (J2000)}& 
\colhead{$\delta$ (J2000)}& 
\colhead{Roll Angle (deg)}& 
\colhead{SWS Speed}} 
\startdata 
N1 & 23 23 29.9 & +58 50 22.8 &  67.26 & 2 \nl 
N2 & 23 23 26.9 & +58 50 22.7 & 234.89 & 2 \nl 
N3 & 23 23 23.9 & +58 50 22.7 & 234.90 & 2 \nl 
N4 & 23 23 20.9 & +58 50 22.7 & 234.91 & 2 \nl 
N5 & 23 23 29.9 & +58 50 52.8 &  67.24 & 1 \nl 
N6 & 23 23 25.7 & +58 50 03.7 &  67.52 & 2 \nl 
S1 & 23 23 38.1 & +58 48 30.9 & 234.87 & 2 \nl 
S2 & 23 23 38.1 & +58 49 00.9 & 234.89 & 2 \nl 
S4 & 23 23 38.1 & +58 48 00.9 & 231.09 & 2 \nl 
S5 & 23 23 35.0 & +58 48 30.8 & 234.90 & 2 \nl 
\enddata 
\end{deluxetable} 
The {\it ISO} data used in this study were obtained using the Short Wavelength 
Spectrometer (SWS) (de Graauw et al. 1996). 
All observations were collected in the AOT 1 
observing mode which provided moderate resolution ($\lambda/\Delta\lambda \sim 
500$) spectra over the entire 
2.38 -- 45.2 $\micron$ wavelength range of the instrument. Ten locations 
in the SNR were targeted as listed in Table 1. The ``North'' fields (N1 -- N6) 
are located in the northern rim of the remnant where many optical knots 
are seen, the X-ray and radio emission are bright, and IR emission is strong. 
Regions N1 -- N4 step across the area of the brightest optical knots. 
Region N5 was intended to sample emission from ejecta at larger radii from 
the expansion center, but was observed for a shorter interval than the other 
regions due to time constraints. Region N6 was aimed at a knot identified by 
Fesen (1990; knot ``D'') as a potential source of [Ne III] emission. 
The ``South'' fields (S1 - S5) are in the southeastern portion of the remnant 
at a location where the 25 $\micron$ emission observed with 
{\it IRAS} and the X-ray emission are fairly strong (Dwek et al. 1987a), 
but the optical and radio emission are relatively weak. 
(Region S3 was not observed due to time constraints.) 
The data have been processed through Off-Line Processing (OLP) 
6.11 and 6.31, with further identification and exclusion of bad 
data and averaging using ISAP\footnote[3]{The ISO Spectral Analysis Package 
(ISAP) is a joint development by the LWS and
SWS Instrument Teams and Data Centers. Contributing institutes are CESR, IAS,
IPAC, MPE, RAL and SRON.} and our own similarly functional software. 
 
Background emission in the spectra is apparently very weak and thus has not
been subtracted for any of the spectra shown here. Some of the South regions 
show little or no line or continuum emission from either the SNR or the 
background. Modeling of the zodiacal light (Kelsall et al. 1998),
the strongest background component at mid-IR wavelengths, indicates
that this emission from local interplanetary dust should contribute less than 
0.25 Jy at 25 $\micron$ in the SWS aperture. 
Similarly, the data have not been corrected for extinction. Using either the 
Mathis (1990) extinction law or that calculated by Drain \& Lee (1984), 
and the visual extinction estimate of $A_V$ = 4.3 mag
(Searle 1971), the largest extinction in the mid-IR should be $\tau \approx 
0.28$, found at the wavelength of the 10 $\micron$ silicate feature. This 
amount of extinction would reduce the observed intensities by $\sim24\%$ at 
$\sim10$ $\micron$. At other wavelengths the extinction generally reduces the 
observed intensities by less than 10 \%. Correction for this would have little 
effect on the results presented here, and would introduce additional sources
of error related to the accuracy of the $V$-band extinction and its 
extrapolation to IR wavelengths, and the uniformity of extinction 
across the remnant.

We had previously obtained spectra of Cas A at three locations using the KAO 
(Moseley et al. 1993). 
These data cover a smaller wavelength range at lower resolution than the 
SWS data. Therefore, their main utility is to provide confirmation of the 
spectral features and relative calibration of the SWS data. The highest quality 
KAO data were obtained at a location roughly matching that of Region N1. The 
other observed locations, at regions S1 and at the optically bright Filament 1 
of Baade \& Minkowski (1954), showed no detectable emission. 
 
\section{ANALYSIS} 
\subsection{Line Emission} 
Selected portions of the SWS spectra of the ten observed fields 
are illustrated in Figure 1. 
The panels in each row of Figure 1, cover a wavelength 
range of exactly $\pm4\%$ 
centered on the rest wavelength of a particular fine structure transition. 
Each row is selected to cover the region of a different transition. In this 
depiction, different transitions observed at the same radial velocity 
in a given field should be aligned in each column. 
\begin{figure}
\epsfysize=6in
\plotone{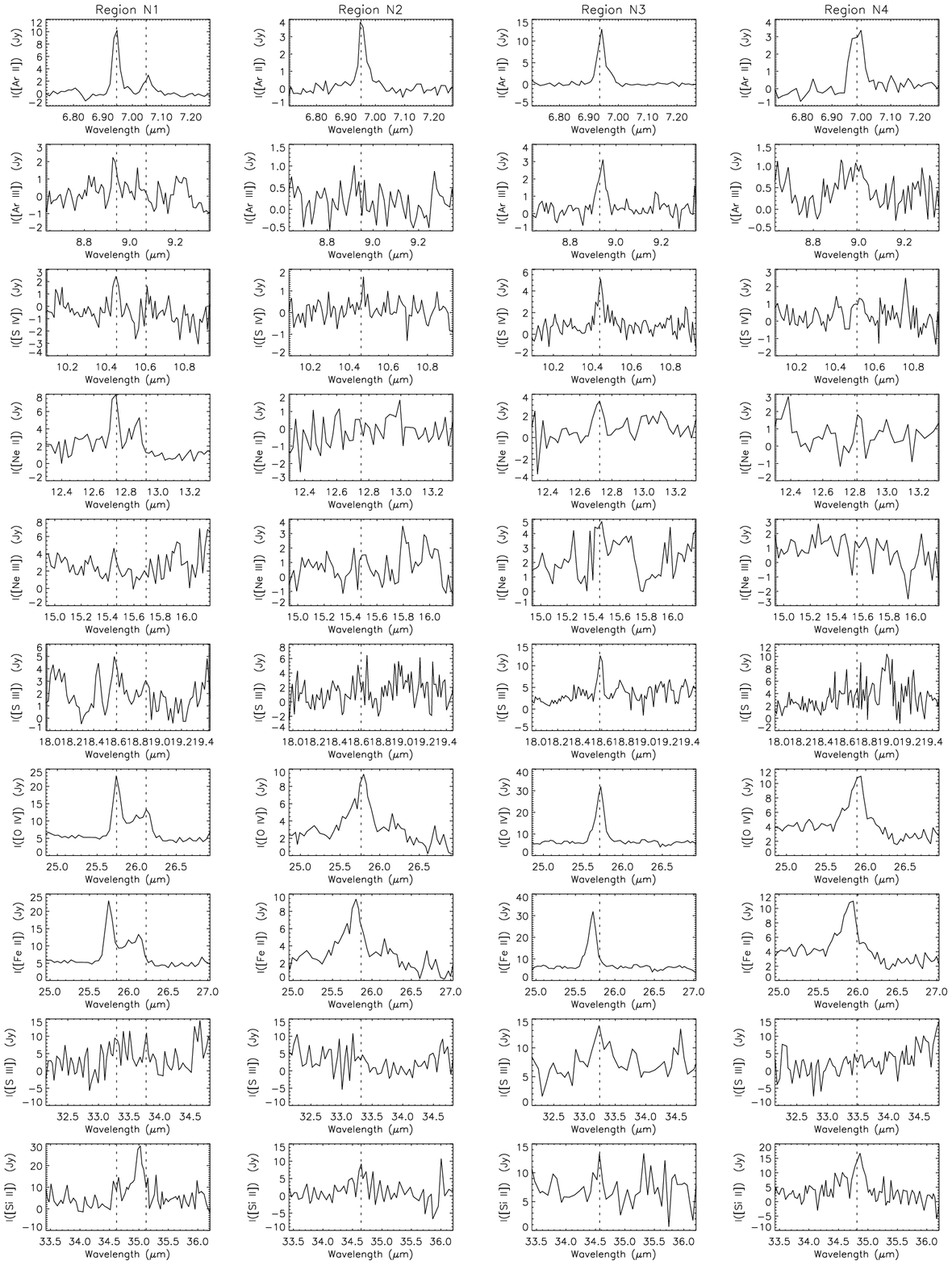}
\caption[]{The spectra of Cas A at the observed locations are 
shown in the regions of selected fine-structure emission lines. 
Each column corresponds to a single location in the Cas A remnant. 
Each row highlights a particular line. The vertical dashed lines 
indicated the expected location of fine structure lines if they are 
found at the same radial velocity as the 26 $\micron$ [O IV] line(s) 
in each observed region. 
\label{Figure 1}} 
\end{figure}

\begin{figure}
\epsfysize=6in
\figurenum{1}
\plotone{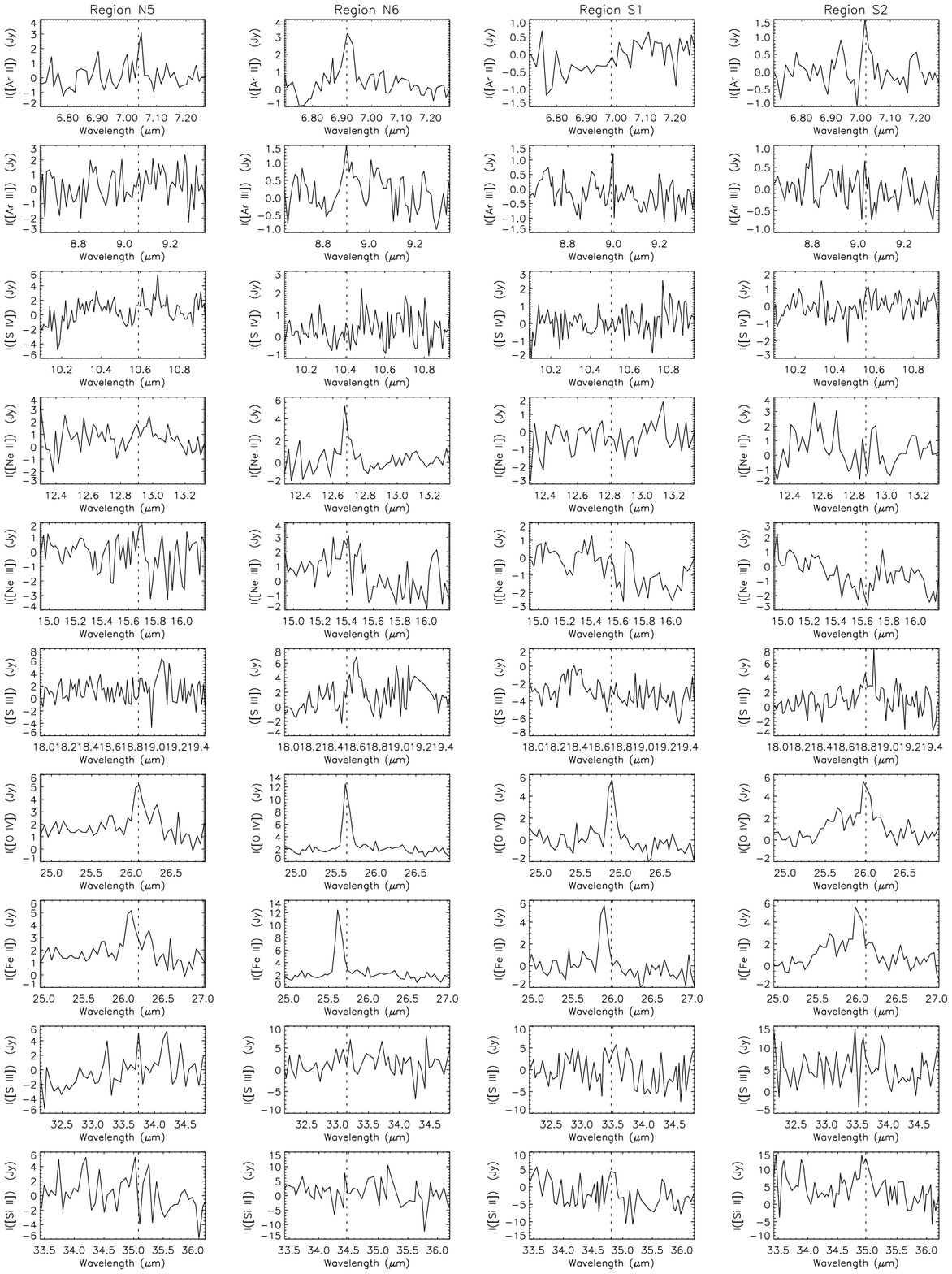}
\caption[]{CONTINUED}
\end{figure}
 
\begin{figure}
\epsfysize=6in
\figurenum{1}
\epsscale{0.5}
\plotone{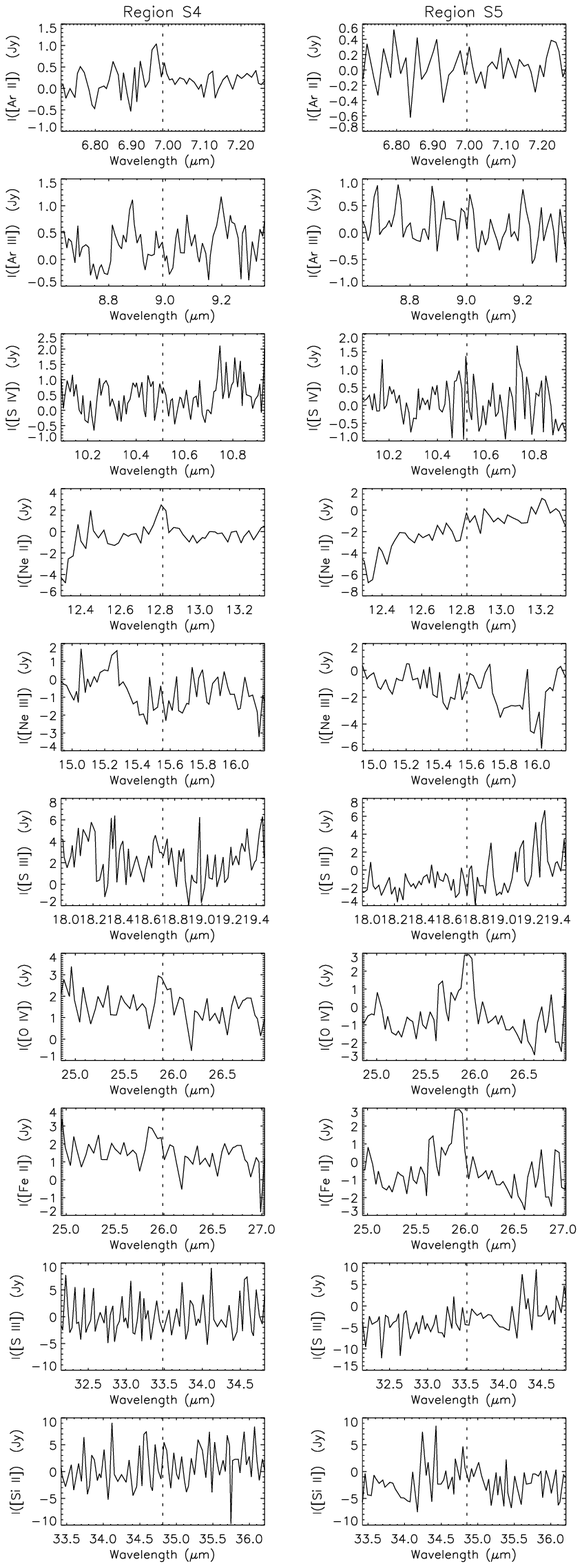}
\caption[]{CONTINUED}
\end{figure}

The most prominent line in the spectra is the [O IV] 25.89 $\micron$ line 
(seventh row of Fig. 1) which is seen in all but the S4 spectrum. 
As well as being intrinsically 
strong, this line falls in a region where the SWS sensitivity is good. 
The dashed lines running vertically in each column of Figure 1 are drawn at 
the location each transition should appear if it occurs at the same radial 
velocity as the [O IV] emission in the field. In Region N1 there are distinct 
red- and blueshifted components. The [Ar II] 6.98 $\micron$ line 
(first row of Fig. 1) is the 
second most prevalent line observed. Like the [O IV] line it benefits from 
being intrinsically strong and in a spectral region where the SWS sensitivity 
is good. The radial velocities of the [Ar II] and the [O IV] lines are 
fairly close. This agreement allows us to discriminate between the 25.89
$\micron$ [O IV] line and the 25.99 $\micron$ [Fe II] line. 
If the 26 $\micron$ line were emitted 
by Fe II, then in Region N1 we would have the unlikely situation where 
the iron ejecta would be moving $\sim10^3$ km s$^{-1}$ faster than the argon 
ejecta on the front side of the SNR, but $\sim10^3$ km s$^{-1}$ {\it slower} 
than the argon on the back side of the remnant. Almost all other lines 
that are observed match the [O IV] and [Ar II] velocities. The exception 
is the apparent [Si II] line in the N1 region. 
All of the detected lines correspond to ionized species that are created by 
overcoming $n$-th ionization potentials in the range of 8 - 55 eV, and that 
have cosmic elemental abundances at least as great as argon. The only species
with similar abundances and ionizations with lines in the SWS range that 
that are not observed are H I and Fe II, Fe III, and Fe V.
 
\begin{deluxetable}{cccccccc}
\tablewidth{0pt} 
\tiny 
\tablecaption{Cas A Line Fluxes\label{Table 2}} 
\tablehead{ 
\colhead{Region}& 
\colhead{Line}& 
\colhead{Rest Wavelength, $\lambda_0$}& 
\colhead{Wavelength, $\lambda$}& 
\colhead{Velocity, $v$}& 
\colhead{Peak Flux Density, $S_{\nu}$}& 
\colhead{Width, $\sigma$}& 
\colhead{Flux, $F$}\\ 
\colhead{}& 
\colhead{}& 
\colhead{($\micron$)}& 
\colhead{($\micron$)}& 
\colhead{(km s$^{-1}$)}& 
\colhead{(Jy)}& 
\colhead{($\micron$)}& 
\colhead{(10$^{-12}$erg s$^{-1}$ cm$^{-2}$)}\\ 
} 
\startdata 
N1     & [Ar II]  & 6.985274 & 6.944 $\pm$ 0.001 & $-1753 \pm$  32 & 11.17 $\pm$ 0.73 & 0.0107 $\pm$ 0.0007 & 18.7 $\pm$ 1.7 \nl 
N1     & [Ar II]  & 6.985274 & 7.052 $\pm$ 0.002 & $+2874 \pm$ 102 & 2.430 $\pm$ 0.28 & 0.0172 $\pm$ 0.0024 & 6.3  $\pm$ 1.1 \nl 
N1     & [Ar III] & 8.99138  & 8.932 $\pm$ 0.002 & $-1988 \pm$  72 & 1.999 $\pm$ 0.90 & 0.0090 $\pm$ 0.0032 & 1.7  $\pm$ 1.0 \nl 
N1     & [S IV]   & 10.5105  & 10.45 $\pm$ 0.005 & $-1842 \pm$ 129 & 2.904 $\pm$ 0.64 & 0.0212 $\pm$ 0.0051 & 4.2  $\pm$ 1.4 \nl 
N1     & [Ne II]  & 12.81355 & 12.73 $\pm$ 0.001 & $-1944 \pm$  28 & 7.613 $\pm$ 0.45 & 0.0212 $\pm$ 0.0013 & 7.5  $\pm$ 0.6 \nl 
N1     & [Ne III] & 15.5551  & 15.45 $\pm$ 0.003 & $-2022 \pm$  58 & 3.813 $\pm$ 0.73 & 0.0116 $\pm$ 0.0034 & 1.4  $\pm$ 0.5 \nl 
N1     & [O IV]   & 25.8903  & 25.75 $\pm$ 0.003 & $-1620 \pm$  37 & 18.10 $\pm$ 1.14 & 0.0449 $\pm$ 0.0030 & 9.2  $\pm$ 0.8 \nl 
N1     & [O IV]   & 25.8903  & 26.12 $\pm$ 0.005 & $+2625 \pm$  61 & 8.649 $\pm$ 0.67 & 0.0591 $\pm$ 0.0058 & 5.6  $\pm$ 0.7 \nl 
N1     & [Si II]  & 34.8152  & 34.99 $\pm$ 0.009 & $+1508 \pm$  79 & 17.24 $\pm$ 1.54 & 0.1156 $\pm$ 0.0092 & 12.2 $\pm$ 1.5 \nl 
\hline 
N2     & [Ar II]  & 6.985274 & 6.956 $\pm$ 0.001 & $-1239 \pm$  58 & 3.952 $\pm$ 0.48 & 0.0112 $\pm$ 0.0016 & 6.9  $\pm$ 1.3 \nl 
N2     & [O IV]   & 25.8903  & 25.77 $\pm$ 0.008 & $-1436 \pm$  94 & 4.917 $\pm$ 0.35 & 0.1109 $\pm$ 0.0086 & 6.2  $\pm$ 0.6 \nl 
\hline 
N3     & [Ar II]  & 6.985274 & 6.941 $\pm$ 0.001 & $-1883 \pm$  32 & 11.46 $\pm$ 0.64 & 0.0133 $\pm$ 0.0006 & 23.8 $\pm$ 1.8 \nl 
N3     & [Ar III] & 8.99138  & 8.938 $\pm$ 0.002 & $-1763 \pm$  71 & 2.949 $\pm$ 0.39 & 0.0137 $\pm$ 0.0021 & 3.8  $\pm$ 0.8 \nl 
N3     & [S IV]   & 10.5105  & 10.44 $\pm$ 0.002 & $-2021 \pm$  62 & 3.869 $\pm$ 0.52 & 0.0145 $\pm$ 0.0021 & 3.9  $\pm$ 0.8 \nl 
N3     & [Ne II]  & 12.81355 & 12.72 $\pm$ 0.005 & $-2230 \pm$ 126 & 3.041 $\pm$ 0.45 & 0.0208 $\pm$ 0.0054 & 2.9  $\pm$ 0.9 \nl 
N3     & [S III]  & 18.7130  & 18.59 $\pm$ 0.002 & $-1998 \pm$  34 & 9.499 $\pm$ 0.76 & 0.0200 $\pm$ 0.0021 & 4.1  $\pm$ 0.5 \nl 
N3     & [O IV]   & 25.8903  & 25.71 $\pm$ 0.002 & $-2100 \pm$  25 & 22.05 $\pm$ 0.99 & 0.0549 $\pm$ 0.0019 & 13.8 $\pm$ 0.8 \nl 
N3     & [S III]  & 33.4810  & 33.28 $\pm$ 0.024 & $-1805 \pm$ 217 & 6.122 $\pm$ 1.40 & 0.1072 $\pm$ 0.0243 & 4.5  $\pm$ 1.4 \nl 
\hline 
N4     & [Ar II]  & 6.985274 & 6.985 $\pm$ 0.002 & $-22   \pm$  67 & 3.643 $\pm$ 0.29 & 0.0196 $\pm$ 0.0015 & 11.0 $\pm$ 1.2 \nl 
N4     & [O IV]   & 25.8903  & 25.89 $\pm$ 0.007 & $-58   \pm$  77 & 6.525 $\pm$ 0.34 & 0.1244 $\pm$ 0.0065 & 9.1  $\pm$ 0.7 \nl 
N4     & [Si II]  & 34.8152  & 34.87 $\pm$ 0.014 & $+454  \pm$ 117 & 14.64 $\pm$ 3.09 & 0.0648 $\pm$ 0.0136 & 5.9  $\pm$ 1.7 \nl 
\hline 
N5     & [O IV]   & 25.8903  & 26.08 $\pm$ 0.011 & $+2230 \pm$ 130 & 3.555 $\pm$ 0.57 & 0.0548 $\pm$ 0.0117 & 2.2  $\pm$ 0.6 \nl 
\hline 
N6     & [Ar II]  & 6.985274 & 6.922 $\pm$ 0.002 & $-2722 \pm$  68 & 3.956 $\pm$ 0.77 & 0.0090 $\pm$ 0.0016 & 5.6  $\pm$ 1.5 \nl 
N6     & [Ar III] & 8.99138  & 8.908 $\pm$ 0.007 & $-2780 \pm$ 223 & 1.224 $\pm$ 0.28 & 0.0248 $\pm$ 0.0078 & 2.9  $\pm$ 1.1 \nl 
N6     & [Ne II]  & 12.81355 & 12.68 $\pm$ 0.003 & $-3150 \pm$  60 & 4.738 $\pm$ 0.46 & 0.0212 $\pm$ 0.0023 & 4.7  $\pm$ 0.7 \nl 
N6     & [O IV]   & 25.8903  & 25.64 $\pm$ 0.002 & $-2949 \pm$  25 & 11.11 $\pm$ 0.85 & 0.0380 $\pm$ 0.0021 & 4.8  $\pm$ 0.5 \nl 
\hline 
S1     & [O IV]   & 25.8903  & 25.88 $\pm$ 0.005 & $-88   \pm$  59 & 6.079 $\pm$ 0.77 & 0.0434 $\pm$ 0.0056 & 3.0  $\pm$ 0.5 \nl 
\hline 
S2     & [O IV]   & 25.8903  & 26.01 $\pm$ 0.007 & $+1385 \pm$  77 & 3.298 $\pm$ 0.37 & 0.0614 $\pm$ 0.0073 & 2.2  $\pm$ 0.4 \nl 
\hline 
S5     & [O IV]   & 25.8903  & 25.91 $\pm$ 0.008 & $+283  \pm$  88 & 3.528 $\pm$ 0.48 & 0.0578 $\pm$ 0.0091 & 2.3  $\pm$ 0.5 \nl 
\enddata 
\end{deluxetable} 
All lines were fit for a baseline and a Gaussian line profile, to determine 
radial velocities, line widths, and fluxes. These results are presented 
in Table 2. In most regions the radial velocities are in excess of 
1000 km s$^{-1}$. 
The velocity uncertainties quoted in Table 2 are the formal uncertainties 
of the fits, and do not account for systematic errors that may influence the 
apparent line positions and widths. 
This indicates that the line emission is arising from 
fast moving ejecta (e.g. the FMKs) and not relatively slow swept up 
material such as the quasistationary flocculi (QSFs) with typical velocities 
of several 10$^2$ km s$^{-1}$. The velocities of the emission can be used to 
identify corresponding knots which have been studied at optical wavelengths. 
Comparison to the observations of Hurford \& Fesen (1996) suggests the 
following correspondence between emission in our observed regions and their 
fast moving knots: Region N1 ($v=-1700$ km$^{-1}$) $\approx$ FMK 2, 
Region N3 $\approx$ FMK 5, and Region N6 $\approx$ FMK 4. The line ratios 
observed in the N regions are similar to those observed with other instruments 
aboard {\it ISO} (Lagage et al. 1996; Tuffs et al. 1998). 
 
The [Ar II] emission is strongly correlated with the [O IV] emission in 
intensity as well as velocity as shown in Figure 2a. Figure 2b shows 
that the ratio of F([Ar II]) / F([O IV]) may increase as a function of 
intensity. However, regions in the southeast portion of the remnant seem 
to exhibit the [O IV] emission without corresponding [Ar II] emission. 
\begin{figure}[t]
\plotfiddle{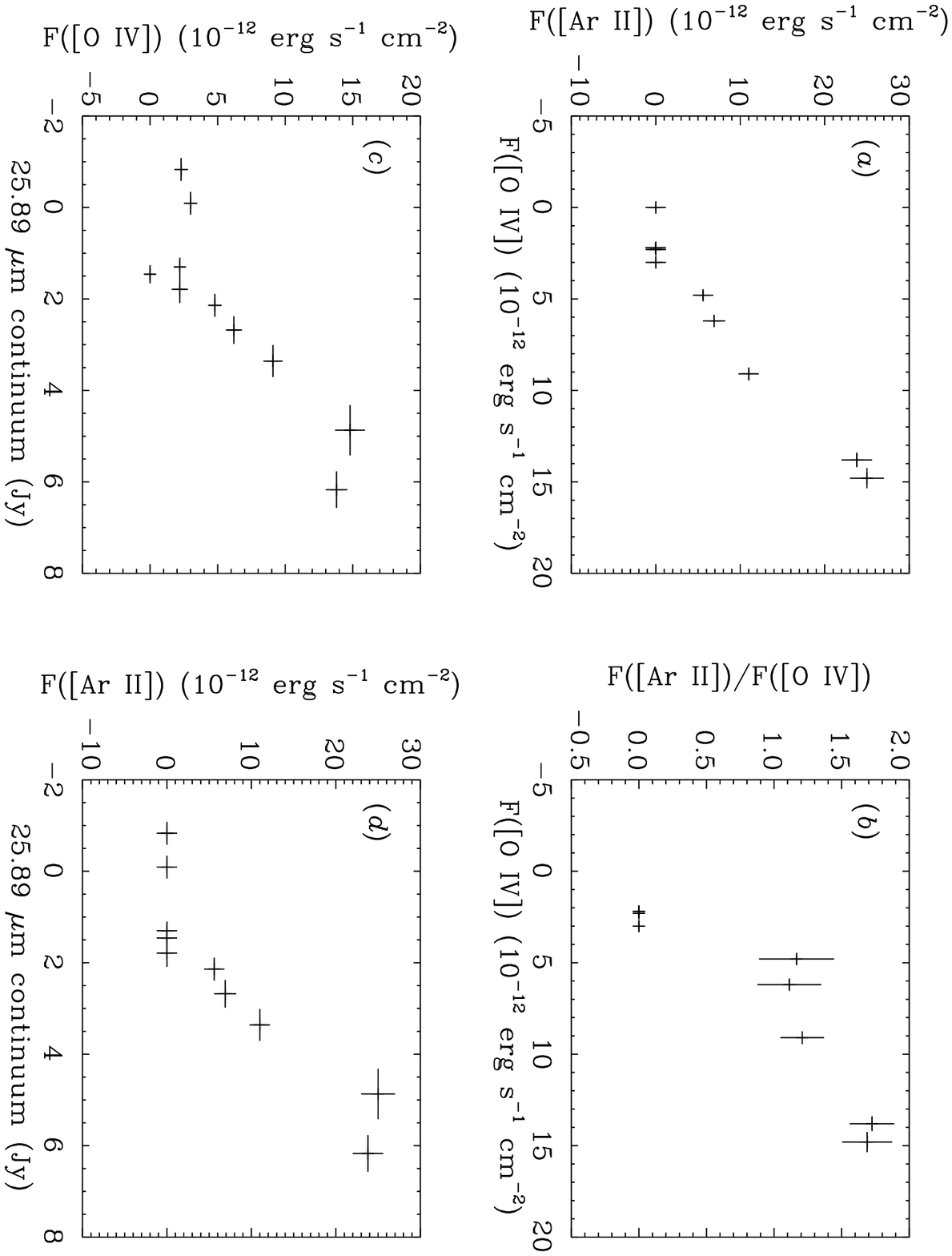}{4in}{90}{50}{50}{180}{0}
\caption[]{Correlations of 
({\it a}) [Ar II] line flux vs. [O IV] line flux, 
({\it b}) [Ar II]/[O IV] flux ratio vs. [O IV] line flux, 
({\it c}) [O IV] line flux vs. continuum flux density, and
({\it d}) [Ar II] line flux vs. continuum flux density, 
\label{Figure 2}} 
\end{figure}
 
The ratio of 18.71 $\micron$ and 33.48 $\micron$ [S III] lines seen in 
Region N3 can be used as a density diagnostic (e.g. Houck et al. 1984). 
The identification and the intensity of the 33.48 $\micron$ line 
is uncertain, but if reliable, the indicated electron density of the emitting 
region is $n_e = 600^{+700}_{-400}$ cm$^{-3}$. Ratios of [Ne III] (15.56 and 
36.01 $\micron$) and [Ar III] (8.99 and 21.83 $\micron$) lines are also 
density sensitive (Butler \& Mendoza 1984; Keenan \& Conlon 1993), 
but for these ionic species one line of each 
pair is too weak to be observed in the SWS data. 
 
Ratios of the [Ar III] 7135\AA\ and 8.99 $\micron$ lines can be used as a 
temperature indicator (Keenan \& Conlon 1993). Using the Hurford \& Fesen 
(1996) intensities for the 7135\AA\ [Ar III] lines, we find electron 
temperatures of $T_e \approx$ 11,000, 4800, and 5500 K for Regions N1, 
N3, and N6 respectively. The temperatures are highly uncertain, and are 
consistently lower than the $T_e \simeq 20,000 - 25,000$ K that Hurford 
\& Fesen (1996) derive from the ratios of the optical [O III] lines. 
This difference 
may be explained if some of the optical emission is missing because the knots 
are wider than the $2\farcs5$ slit used for the optical spectroscopy, or 
if the IR emission arises from a blend of several knots. Alternatively, it 
may be that the [Ar III] and [O III] are tracing different regions of the 
shocked FMK. 

Estimates of the mass in each of the ionic species observed have been made 
assuming that the electron temperature is $T_e = 10^4$ K and the 
electron density is $n_e = 1000$ cm$^{-3}$, which is at least 
an order of magnitude below the critical densities at which the observed IR 
transitions would be collisionally deexcited (Table 3). 
At densities below the critical densities, the derived masses are  
proportional to $T_e^{0.5} e^{hc/\lambda kT_e} n_e^{-1}$.
\begin{deluxetable}{lccc}
\tablewidth{0pt} 
\tiny
\tablecaption{Derived Masses in Bright Regions\label{Table 3}} 
\tablehead{ 
\colhead{Ionic}& 
\colhead{}& 
\colhead{Mass\tablenotemark{a} \ (10$^{-5}$ M$_{\sun}$)}& 
\colhead{}\\\cline{2-4} 
\colhead{Species}& 
\colhead{N1 ($v<0$)}& 
\colhead{N1 ($v>0$)}& 
\colhead{N3} 
} 
\startdata 
O IV   & 2.3     &  1.4                  & 3.4 \nl 
Ne II  & 19      & \nodata               & 7.5 \nl 
Ne III & 2.7     & \nodata               & \nodata \nl 
Si II  & \nodata &  2.9\tablenotemark{b} & \nodata \nl 
S III  & \nodata & \nodata               & 1.1 \nl 
S IV   & 0.34    & \nodata               & 0.32 \nl 
Ar II  & 27      &  9.2                  & 35 \nl 
Ar III & 1.3     & \nodata               & 2.8 \nl 
\hline 
Dust   & \nodata & \nodata               & 0.28 \nl 
\enddata 
\tablenotetext{a}{Ionic masses assume $n_e = 1000$ cm$^{-3}$,
$T_e = 10^4$~K, and that collisional deexcitation is unimportant.} 
\tablenotetext{b}{Velocity differs from [O IV] and [Ar II] emission} 
\end{deluxetable} 
 
\subsection{Continuum Emission} 
\subsubsection{Dust Composition and Mass} 
The IR spectra from the various regions also exhibit continuum emission with 
an intensity that varies as widely as the line intensities. 
While some fraction of the continuum emission of the SNR should be synchrotron 
radiation, the bright continuum emission that is detected in the SWS data
is far too strong to be an extension of the radio power-law spectrum through
the IR wavelengths (e.g. Mezger et al. 1986; Dwek et al. 1987a). The continuum
emission also exhibits a shape that strongly indicates that it is 
thermal emission from dust grains within the SNR.

We find that the continuum flux density at 26 
$\micron$ (the baseline flux density underlying the fit to the [O IV] 
lines) is well correlated with the 
line intensities of both [O IV] and [Ar II] (Figures 2c and 2d). This suggests 
that the dust that produces the observed continuum emission is associated 
with the same fast moving ejecta which produces the line emission. A similar 
correlation between the [S IV] 10.5 $\micron$ emission and the 11.3 $\micron$ 
continuum has led Lagage et al. (1996) to the same conclusion. 
 
In the following, we will concentrate on the analysis of the region N3, which 
was the brightest of the regions and has the cleanest spectrum. Other than 
the changes in intensity, it is not clear that any of the other continuum 
spectra are different from that of region N3. The continuum spectrum after 
clipping the observed spectral lines and averaging to a resolution of 
$R = \lambda / \Delta\lambda = 200$ is shown 
in Figure 3. Also shown is the KAO spectrum, arbitrarily scaled to match. 
The KAO and {\it ISO} results appear to be in very good agreement. 
Both data sets 
clearly indicate a spectrum that rises quickly, peaks at 21 -- 22 $\micron$ 
and then fades more slowly at longer wavelengths. 
The {\it IRAS} flux densities at 12, 25, 60, and 
100 $\micron$ normalized to roughly match the 25 $\micron$ flux density of 
Region N3 are shown in Fig. 3. 
\begin{figure}[t]
\epsscale{0.5}
\plotfiddle{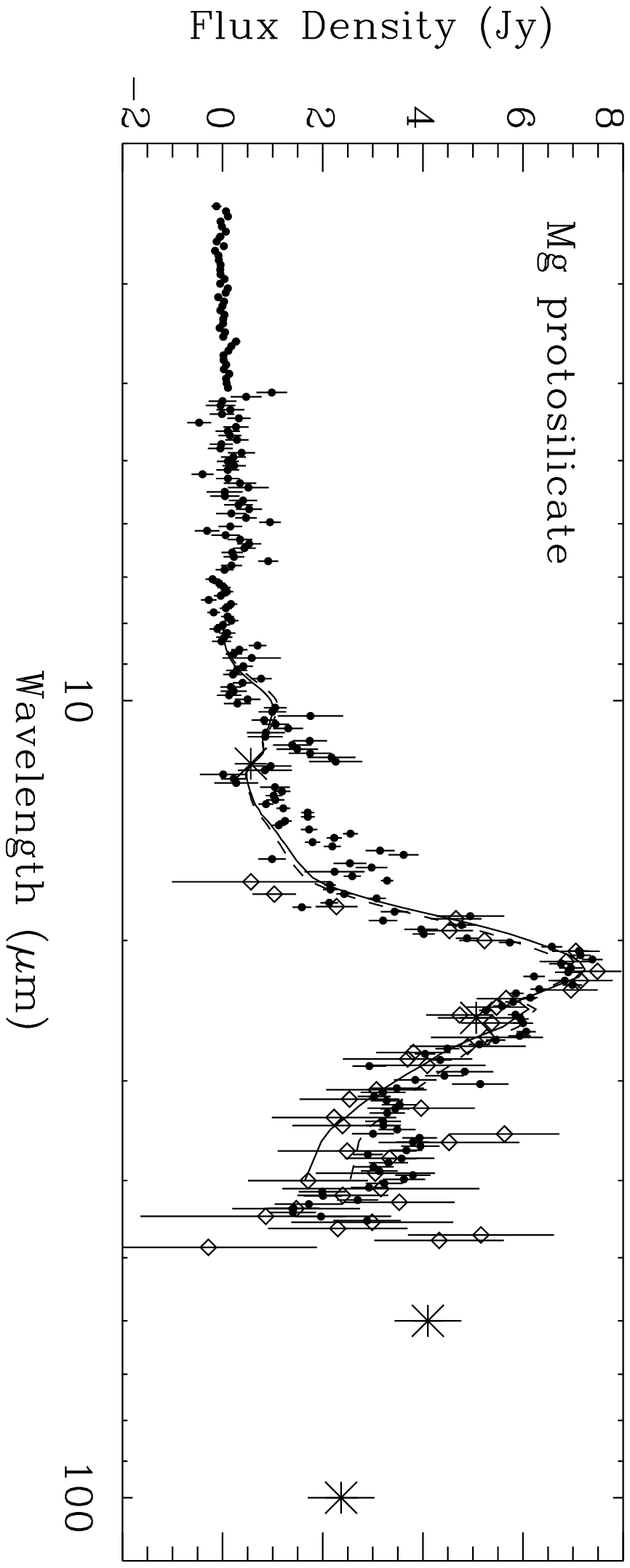}{3in}{90}{75}{75}{288}{0}
\vspace{-0.5in}
\caption[]{The {\it ISO} SWS spectrum for the 
N3 region (dots with 1 $\sigma$ 
error bars) smoothed to a resolution of $\lambda$/$\Delta\lambda$ = 200. 
The KAO data (normalized to the SWS data) are indicated by diamonds. The 
solid line is the best fit spectrum for dust with a Mg protosilicate 
emissivity (Dorschner et al. 1980; sample F), for which $T_d = 169$ K. 
The dashed line is the model spectrum if the dust is stochastically heated
(\S3.2.3). The broadband {\it IRAS} data at 12, 25, 60 and 100 
$\micron$ integrated over the 
entire SNR are scaled by a factor of 40 and plotted as stars. 
\label{Figure 3}} 
\end{figure}

The peak in this spectrum 
cannot be fit with single temperature blackbody emission from dust with a 
$\lambda^{-2}$ emissivity. 
The line drawn through Figure 3 indicates a single dust temperature fit 
to the SWS spectrum for dust grains with an emissivity taken to be that 
of Mg protosilicate (Dorschner et al. 1980). The derived dust temperature for 
this fit is 169 K, and can vary by $\sim40$ K without producing a distinctly 
poor fit to the data. The best fitting spectra for other potential grain 
compositions are illustrated in Figure 4. Blackbody emission, graphite, and 
silicate grains with Draine \& Lee (1984) optical constants provide 
relatively poor fits. The 
22 $\micron$ peak in the Cas A spectrum is at a wavelength too long 
for the typical astronomical silicate, which is usually stated to have an 
emission feature at $\sim18~\micron$. Of the various silicate optical 
properties which have been measured and published, the only ones which produced 
good fits to the Cas A spectrum were the protosilicates measured by Dorschner 
et al. (1980). For the SWS data at region N3, the Fe protosilicate emissivity 
provides a slightly better fit than the Mg protosilicate. 
Considering only the KAO data, 
the Mg protosilicate provides the better fit. Ca protosilicate has a emission 
feature at 22 $\micron$, but it is narrower than that of the other 
protosilicates and the data. Iron oxide (FeO) has also has a sharp emission 
feature near 22 $\micron$ (Henning et al. 1995). 
The feature can be broadened to produce a 
fair fit to the data if it is assumed that, rather than spherical grains, a 
continuous distribution of ellipsoidal grains (CDE approximation; 
Bohren \& Huffman, 1983) is present. However, FeO grains cannot account for 
both the $22 \micron$ feature and the continuum emission at $>30~\micron$. 
\begin{figure}[t]
\epsscale{0.5}
\plotone{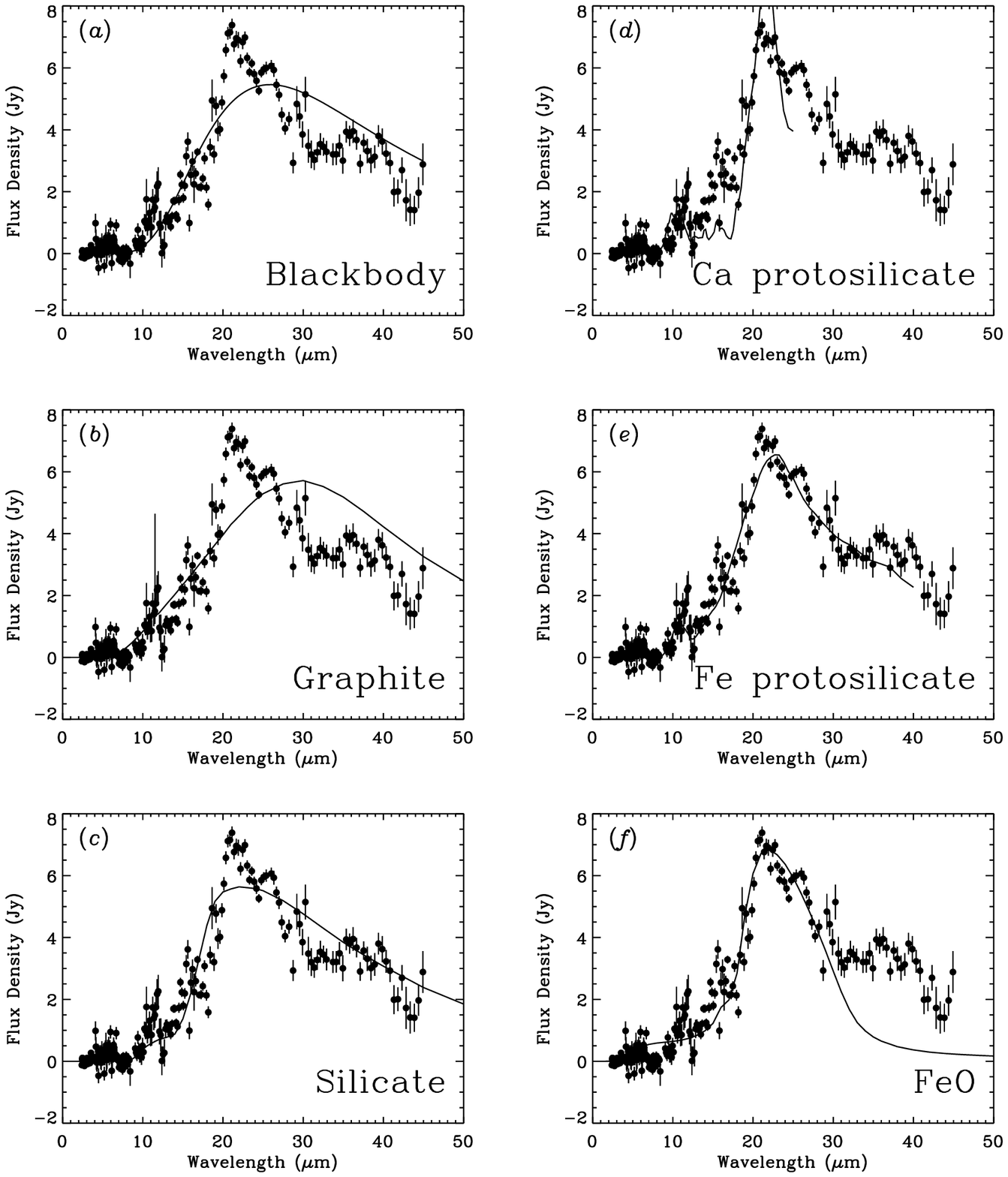}
\caption[]{The same SWS data for Region N3 as in Figure 3, this time fitted by 
({\it a}) a 112 K blackbody spectrum modified by a $\lambda^{-2}$ emissivity, 
({\it b}) a 188 K graphite spectrum (Draine \& Lee 1984), 
({\it c}) a 124 K astronomical silicate spectrum (Draine \& Lee 1984), 
({\it d}) a 184 K Ca protosilicate spectrum (Dorschner et al. 1980; sample A), 
({\it e}) a 165 K Fe protosilicate spectrum (Dorschner et al. 1980; sample E), and
({\it f}) a 396 K FeO spectrum calculated for a continuous distribution of 
ellipsoidal grains (Henning et al. 1995). 
\label{Figure 4}} 
\end{figure}
 
The {\it IRAS} measurements shown in Figure 3, show an excess at 60 and 100 
$\micron$ over the extrapolated spectra that fit 
the 22 $\micron$ emission peak (Figs. 3 or 4{\it e,f}). This 
indicates the presence of cooler dust within the SNR. However, since the 
{\it IRAS} flux densities are for the integrated emission of the SNR, it is 
not clear whether the cooler dust resides in the FMKs or elsewhere in the SNR. 
 
The mass of hot ($\sim170$ K) dust observed in Cas A is given by 
\begin{equation} 
M_{dust} = \frac{S_{\nu}\ d^2}{B_{\nu}(T_d)\ \kappa_{\nu}} 
\end{equation} 
where $S_{\nu}$ is the observed flux density, $d$ is the distance to Cas A, 
$B_{\nu}(T_d)$ is the Planck blackbody function evaluated at the dust 
temperature $T_d$, and $\kappa_{\nu} = 3 Q/4 a \rho$ is the mass absorption 
coefficient for the dust. For region N3, $S_{\nu}(26\ \micron) = 6.2$ Jy, 
$d = 3.4$ kpc (Reed et al. 1995), 
$T_d = 169$ K, and $\kappa_{\nu}(26\ \micron) = 1332$ cm$^2$ 
g$^{-1}$ for Mg protosilicate (Dorschner et al. 1980), yielding $2.8 
\times 10^{-6}$ M$_{\sun}$ of dust in the region. Comparison to the mass of 
the gas in Region N3 (Table 3) suggests that as much as 2\% of the Si ejecta 
in the FMKs has been condensed into dust assuming that Mg protosilicate is 
$\sim25\%$ Si by mass, and all the Si is traced by the observed dust 
and [Si II] emission. 
 
The total mass of hot dust 
in the SNR can be roughly estimated by scaling this dust mass by the ratio 
of the total flux density of Cas A to the flux density of this region. With 
a total flux density from {\it IRAS} observations of $\sim164$ Jy (Arendt 1989) 
at 25 $\micron$, we estimate that Cas A contains $\sim7.7 \times 10^{-5}$ 
M$_{\sun}$ of hot dust. The mass derived here is a factor of 
10 -- 100 lower than 
previous estimates of $0.5 - 7\times 10^{-3}$ M$_{\sun}$ based on {\it IRAS} 
observations (Mezger et al. 1986; Braun 1987; Dwek et al. 1987a; 
Greidanus \& Strom, 1991; Saken, Fesen \& Shull 1992). 
There are two reasons why our dust mass estimate is lower. First, the mass 
absorption coefficient for Mg protosilicate is about a factor of 2 larger than 
that applied in the other studies. Second and more importantly, the dust 
temperature we find is 
significantly higher than those estimated from the {\it IRAS} observations. 
Part of this difference is the high temperature implied when fitting the data 
with Mg protosilicate emissivity (cf. Figs. 3 and 4{\it c}), and part of the 
difference is that the SWS data do not include the 60 and 100 $\micron$ 
and are therefore insensitive to the presence of cooler dust grains in the 
FMKs or elsewhere in the SNR. 

The apparent excess of 60 and 100 $\micron$ emission in Figure 3 suggests the
presence of a second, cooler dust component in the Cas A SNR. 
A fit to the 60 and 100 $\micron$ flux densities only gives a dust temperature
of 52 K (Arendt 1989), and a resulting cool dust mass of 
$\sim3.8 \times 10^{-2}$ M$_{\sun}$ for the entire SNR. (Note 
that masses quoted by Arendt 1989 are too large by a factor of $\pi$.) The
25 $\micron$ emission of this cooler dust would be $\sim 18$ Jy or $\sim10\%$
of the total emission. This contribution would not greatly alter the 
spectral fitting discussed previously.
 
\subsubsection{Dust Heating and Location} 
The collisional heating rate for a dust grain of radius $a$ (in cm) is 
given by (Dwek 1987): 
\begin{equation} 
H_{coll}(a,T_e) = \pi a^2 \left(\frac{32}{\pi m_e}\right)^{1/2} 
n_e (kT_e)^{3/2} h(a,T_e) 
\end{equation} 
where $m_e$, $n_e$, and $T_e$ are the electron mass, number density, and 
temperature, and $h(a,T_e)$ is a grain heating efficiency, which approaches 
a value of $\sim 1.0$ for larger grains and lower temperatures. 
The radiative heating rate of the same dust grain can be expressed as: 
\begin{equation} 
H_{rad}(a) = \pi a^2 \langle Q_{\nu}\rangle c U 
\end{equation} 
where $U$ is the radiative energy density and $\langle Q_{\nu}\rangle$ denotes 
the spectrally averaged absorption coefficient. 
In the region immediately behind a shock passing through a FMK, we can 
approximate the energy density as $U = \frac{1}{c}\int I d\omega = \frac{1}{c} 
l n_e^2 \Lambda$ where $l$ is a mean scale length of the emitting region (e.g. 
the cooling length of the shock), and 
$\Lambda$ is the cooling function of the shocked gas 
in units of erg cm$^3$ s$^{-1}$. 
The ratio of these heating rates is thus
\begin{equation}
\frac{H_{coll}}{H_{rad}} = \left(\frac{32 k T_e}{m_e c^2}\right)^{1/2} 
                \frac{n_e k T_e }{U} \frac{h(a,T_e)}{\langle Q_{\nu}\rangle}
\end{equation}
The sum of these heating rates will be equal to the grain cooling rate or 
luminosity: 
\begin{equation} 
L_{gr} = 4 \pi a^2 \sigma T_d^4 \langle Q(a,T_d)\rangle 
\end{equation} 
where $T_d$ is the dust temperature, and 
$\langle Q(a,T_d)\rangle$ is the dust absorption coefficient averaged 
over a Planck spectrum of temperate $T_d$.
 
Equating the grain heating and cooling rates, 
we can solve for the grain temperature 
given estimates to the gas temperature, density, and cooling function. 
Sutherland \& Dopita (1995) present models of the structure and emission from 
shocked knots of O-rich ejecta. Hurford \& Fesen (1996) 
suggest that these models with cloud shock velocities of $\sim 
150-200$ km s$^{-1}$ may be appropriate for the Cas A FMKs, if the extinction 
is $\sim 1$ mag greater than the commonly adopted $A_V = 4.3$ mag 
(Searle 1971). 
Thus, the 200 km s$^{-1}$ shock model of Sutherland \& Dopita (1995) estimates 
a postshock temperature of $T_e = 10^{6.64}$ K, $n_e = 400$ cm$^{-3}$, and 
$\Lambda = 10^{-17.5}$ erg cm$^3$ s$^{-1}$. For these conditions, collisional 
heating dominates radiative heating by factors of several hundred, and the 
dust temperature is given by 
\begin{equation} 
T_d = \left[\frac{\left(\frac{32}{\pi m_e}\right)^{1/2} 
      n_e (kT_e)^{3/2} h(a,T_e)} {4 \sigma \langle Q \rangle}\right]^{1/4}. 
\end{equation} 
For silicate grains with $T_d > 150$ K, $\langle Q \rangle = 0.25\ 
(a/1\ \micron)$ (Draine \& Lee 1984).
Approximating $h \approx 1.0$, the estimated temperature 
for $a$ = 0.01 $\micron$ grains will be $T_d \approx 180$ K. Larger grains will 
be correspondingly cooler. This simple calculation shows that 
the high dust temperature indicated by the continuum spectrum of Cas A is 
consistent with emission from collisionally heated dust in the post shock 
region of FMKs, but not with radiatively heated dust in the FMKs. 
Note that the collisional heating mechanism is only tenable if 
the electron temperature equilibrates with the ion temperature as in the models 
presented by Sutherland \& and Dopita (1995), in contrast to some of the 
pure oxygen shock models calculated by Itoh (1988) where electron temperatures 
only reach $T_e\approx10^5$ K. 
 
In Region N3, the integrated flux of thermal emission from the hot dust is 
$\sim3$ times the total flux in IR, optical, and UV lines, assuming that 
Hurford \& Fesen (1996) 
FMK5 corresponds to the observed knot, and that the spectrum of Sutherland \& 
Dopita's OSP200 is a reasonable estimate of the UV line intensities. This 
suggests that thermal radiation from collisionally heated dust is the most 
important cooling mechanism for the shocked FMKs, in agreement with the more 
general comparison of the relative importance of X-ray and IR cooling found
by Dwek et al. (1987b) for various SNRs.

While the hot dust appears to be associated with the FMKs, the location of 
the cooler dust producing the 60 and 100 $\micron$ emission is not as clear.
If the cooler dust component is also associated with the FMKs, it could be 
dust within the unshocked (or cooled post-shock) volume of the FMKs. In 
a cooler environment, collisional heating of the dust may no longer be 
significant, but the radiation of the shock passing through a typical FMK
may be sufficient to heat nearby dust to the required temperature of 
$T_d \approx 50$ K. In this case, the ratio of the mass of hot dust 
to the mass of cool dust ($\sim1/500$) will be equal to the fraction of the 
FMK mass that is contained within the hot post-shock region. 
However, the global 60 and 100 $\micron$ emission depicted in Fig. 3 may not
originate in the FMKs. It may be emitted by dust in the 
X-ray emitting gas between and around the FMKs. 
A dust temperature of $\sim50$ K 
would then imply electron densities of $\sim 1$ cm$^{-3}$ if the
dust grains have radii of $a = 0.1$ $\micron$ (Dwek 1987). Higher 
electron densities would be required if the grains are larger. This derived 
density is not atypical for SNRs, but is somewhat low compared to some
recent estimates of the density, $n_et \sim 100\times 10^{11}$ cm$^{-3}$ s, 
derived from the X-ray observations of Cas A 
(Holt et al. 1994; Vink, Kaastra, \& Bleeker 1996; Favata et al. 1997). 
 
\subsubsection{A Stochastic Heating Alternative} 
In the previous sections, the correlation of the continuum emission with 
the line emission of the FMKs is taken as evidence for dust radiating 
at equilibrium temperatures of $\sim$ 170 K and collisionally heated by the
shocked FMK gas. The composition of the dust appears 
to reinforce this conclusion. However, an alternate model for the strength 
and morphology of the continuum emission can be constructed, assuming that 
the bulk of the dust is embedded in within the X-ray emitting gas rather 
than the FMKs. In this case, as {\it IRAS} observations had indicated, 
most of the dust is collisionally heated to temperatures of $\sim100$ K in the 
hot X-ray emitting portion of the SNR. Then in regions where the FMKs are 
plowing into the reverse shock, the bowshocks of the knots 
compress the gas which will increase the dust temperatures and 
the spatial density of dust grains (Sutherland \& Dopita 1995). However, 
this environment is at a significantly lower density than the material of the
FMKs. Therefore, if the dust grains are small enough they will be 
stochastically heated, i.e. the grain cooling time will be shorter than the 
collision time, and the grain temperatures will fluctuate instead of remaining 
at equilibrium temperatures. From the work of
Dwek (1986), we see that silicate grains with radii of $a = 0.005$ $\micron$
immersed in a gas with $n_e = 10$ cm$^{-1}$ and $T_e = 10^7$ K will have 
dust temperatures $30 < T_{dust} < 200$ K. 
The presence of the colder grains leads to additional emission at longer 
wavelengths. For the example chosen here, this improves the fit between the 
model and the SWS and KAO data (dashed line in Fig. 3).

While dust heated in the bow shocks of the FMKs may be responsible for the
correlation of the optical FMKs with the brightest emission in the 11.3
$\micron$ image (Lagage et al. 1996), the fainter diffuse emission that forms
a nearly complete shell correlates fairly well with both X-ray and radio
emission (e.g. Anderson \& Rudnick 1995). This shell would be emission from
stochastically heated dust grains in the lower density X-ray emitting gas. In
addition, the shell could contain significant synchrotron emission, depending
on the actual shape of the extension of the radio synchrotron spectrum into
the IR regime (e.g. Mezger et al. 1986; Dwek et al. 1987a; Tuffs et al. 1998).
The morphology of the X-ray and radio emission are similar after correction
for extinction (Keohane, Rudnick \& Anderson 1996; Keohane, Gotthelf, \& Petre
1998), making it difficult to distinguish thermal from synchrotron emission
by the apparent structure.

\section{DISCUSSION} 
\subsection{Mixing of the Ejecta} 
On a large scale, the Cas A SNR gives the impression of a star that has been 
dissected by layers. First, most of the H- and N-rich outer layers are lost in 
winds prior to the SN explosion, leading to the QSFs 
(van den Bergh 1971; Peimbert \& van den Bergh 1971). 
Then, in the SN explosion, the outermost layer of the star 
is blasted away at the highest velocity, producing the N-rich FMFs 
(Fesen, Becker, \& Blair 1987; Fesen, Becker, \& Goodrich 1988). 
Inner layers of the star produce the FMKs, some of which are 
mainly oxygen, and others, from deeper layers 
of the star, contaminated with O-burning products (Si, Ar, Ca) 
(Chevalier \& Kirshner 1979; Winkler, Roberts, \& Kirshner 1991). 
All of these knots of ejecta are rendered visible when they are 
shocked in the expanding shell of the SNR blast wave. The QSFs are caught 
by the shell, at the forward shock; the fast-moving knots and flocculi 
catch the decelerating shell from the inside, at the reverse shock. 
 
Within this framework, the lack of [Ar II] emission from the 
South regions (\S 3.1) suggests 
that the material in this region originated in layers of the star above the 
O-burning zone. The observed ejecta is presumably trailed by ejecta enriched 
with O-burning products which has yet to catch up with the reverse shock 
because either the knots have lower velocities or the 
reverse shock is located at a larger radius
than in the northern portion of the SNR. 
Additional suggestions of a lack of mixing in the SN ejecta are provided
by the composition of the dust, discussed below.
 
Despite the large scale differentiation of the ejecta, evidence for partial 
mixing of the ejecta is provided by the observation of [Fe II] and [Ni II] 
in optical spectra of FMKs (e.g. Winkler, Roberts, \& Kirshner 1991), 
and now by the detection of IR lines of both Ne and Ar in the FMK spectra 
in the North regions of the SNR (Lagage et al. 1996; Tuffs et al. 1998). 
The argon, iron, and nickel originate in layers of the progenitor that 
lie below the oxygen-burning shell, whereas the neon is found in layers above
the oxygen-burning shell 
(Weaver, Zimmerman \& Woosley 1978; Woosley, Axelrod, \& Weaver 1984; 
Johnston \& Yahil 1984). 
 
\subsection{Dust Formation in the Ejecta} 
Since the Cas A SNR is still sweeping up the circumstellar material 
shed by its progenitor, the dust in the SNR must be relatively recently formed 
either in the pre-SN stellar wind, or from the ejecta of the SN itself. 
The correlation of the continuum intensity with the O and Ar line strengths 
of the FMKs suggest that the observed dust has formed from the ejecta. 
In \S 3.2, we found that the continuum spectrum is well represented by 
dust with optical properties similar to those the Mg protosilicates 
measured in the lab by Dorschner et al. (1980). We favor the Mg over the Ca 
and Fe protosilicates because of the good match to the 
position and width of the 22 $\micron$ emission feature, and because 
the Mg abundance should 
exceed the Ca and Fe abundances in the O-rich layers of evolved massive stars 
(Weaver et al. 1978; Woosley et al. 1984), with no mixing of the ejecta 
required. 
Similar mixing considerations have led Clayton et al. (1997) to propose
that type X SiC grains found in meteorites must originate in type
Ia SNe, rather than type Ib or II SNe as Cas A is generally believed to be.
We also favor the Mg protosilicate 
identification of the dust over the possibility of FeO dust, because 
while FeO dust could create the 22 $\micron$ feature, its
spectrum drops quickly at $\lambda > 25$ $\micron$ and would
still require another dust component to provide the longer wavelength emission
seen in the SWS spectra and in the {\it IRAS} data. Finally, the 400 K dust
temperatures required for the FeO grains is higher that we expect to find
in the SNR at this time.

The apparent presence of Mg protosilicates does not preclude the presence
of carbonaceous dust which may have formed in the ejecta. 
The fate of C in the Cas A ejecta is difficult to determine. Normally strong
C lines are obscured by high extinction at UV wavelengths, 
optical lines are weak (Hurford \& 
Fesen 1996), and no C lines are present in the spectral range of the SWS.
If the carbon is locked up in graphitic dust, its IR spectrum will lack 
any distinctive features in contrast to the strong spectral 
features of silicate grains.
The composition of the dust required to explain the 60 and 100 $\micron$
emission seen by the IRAS is unconstrained by the broadband observations. 
This dust
component could be either additional silicate dust or carbonaceous dust
since only a fraction of the carbon in the ejecta is expected to be locked
up in CO molecules (Clayton, Liu, \& Dalgarno 1998).

\section{Conclusion}
The {\it ISO} SWS data have allowed a detailed look at the line and continuum 
emission from the Cas A SNR. The data provide new information on the 
composition of both the gas and dust in the supernova ejecta. The line 
emission is dominated by [O IV] 25.89 $\micron$, which is 
not surprising for this prototype of
oxygen-rich SNRs. Emission from oxygen-burning products, particularly [Ar II] 
7.99 $\micron$, is also strong in some regions of the SNR. 
The dust composition revealed by these data clearly show that the dust formed
in the ejecta differs from typical interstellar silicates. A Mg protosilicate
composition is suggested for the recently formed dust. 
The mass of the hot dust observed appears to be a 
small fraction of the condensible silicon. However, a cooler dust component is
also present in the SNR, though not necessarily associated with the FMKs.

Somewhat unexpectedly, the SWS data do not provide evidence for emission 
from iron in the SN ejecta, in either the gas phase or the dust.
Another unresolved question concerns the apparent morphological differences
that are implied between the $20-25$ $\micron$ emission observed with the SWS
and previous {\it IRAS} and ground-based studies
(Dwek et al. 1987a; Greidanus \& Strom 1991). 
 
\acknowledgements 
We wish to thank A. Noriega-Crespo, S. Unger, and the IPAC staff for help 
with data reduction. R. Smith, W. Glaccum, and K.-W. Chan provided useful 
assistance with the KAO data and its interpretation. We thank 
H. Mutschke and the group at Friedrich-Schiller-Universit\"at 
Jena for making their optical data available on the internet. This analysis 
was funded through NASA's ISO Guest Observer program.
 


\begin{references} 
\reference{And95} Anderson, M. C., \& Rudnick, L. 1995, \apj, 441, 307
 
\reference{Are89} Arendt, R. G. 1989, \apjs, 70, 181 
 
\reference{Baa54} Baade, W., \& Minkowski, R. 1954, \apj, 119, 206 
 
\reference{Boh83} Bohren, C. F., \& Huffman, D. R. 1983, Absorption and 
Scattering of Light by Small Particles, (New York: Wiley) 
 
\reference{Bra87} Braun, R. 1987, \aap, 171, 233 
 
\reference{But84} Butler, K., \& Mendoza, C. 1984, \mnras, 208, 17P 
 
\reference{Che79} Chevalier, R. A., \& Kirshner, R. P. 1979, \apj, 233, 154 
 
\reference{Cla97} Clayton, D. D., Arnett, D., Kane, J., \& Meyer, B. S. 1997, 
\apj, 486, 824

\reference{Cla98} Clayton, D. D., Liu, W., \& Dalgarno, A. 1998, Science, 
in press

\reference{deG96} de Graauw, T., et al. 1996, \aap, 315, L49 
 
\reference{Din82} Dinerstein, H. L., Capps, R. W., Dwek, E., \& Werner, M. W. 
1982, \apj, 255, 552 
 
\reference{Din87} Dinerstein, H. L., Lester, D. F., Rank, D. M., Werner, M. W., 
\& Wooden, D., H. 1987, \apj, 312, 314 
 
\reference{Dor80} Dorschner, J., Friedmann, C., Guertler, J., \& Duley, W. W. 
1980, \apss, 68 159 
 
\reference{Dra84} Draine, B. T., \& Lee, H. M. 1984, \apj, 285, 89 
 
\reference{Dwe86} Dwek, E. 1986, \apj, 302, 363 

\reference{Dwe87} Dwek, E. 1987, \apj, 322, 812 

\reference{Dwe92} Dwek, E., \& Arendt, R. G. 1992, \araa, 30, 11
 
\reference{Dwe87a} Dwek, E., Dinerstein, H. L., Gillett, F. C., 
Hauser, M. G., \& Rice, W. L. 1987a, \apj, 315, 571 
 
\reference{Dwe87b} Dwek, E., Petre, R., Szymkowiak, A., \& Rice, W. L. 1987b, 
\apj, 320, L27 

\reference{Dwe81} Dwek, E., \& Werner, M. W. 1981, \apj, 248, 138

\reference{Fav97} Favata, F., Vink, J., Dal Fiume, D., Parmar, A. N., 
Santangelo, A., Mineo, T., Preite-Martinez, A., Kaastra, J. S., \& Bleeker, 
J. A. M. 1997, \aap, 324, L49
 
\reference{Fes87} Fesen, R. A., Becker, R. H., \& Blair, W. P. 1987, \apj, 
313, 378 
 
\reference{Fes88} Fesen, R. A., Becker, R. H., \& Goodrich, R. W. 1988, \apj, 
329, L89 
 
\reference{Fes90} Fesen, R. A. 1990, \aj, 99, 1904 
 
\reference{Gre91} Greidanus, H., \& Strom, R. G. 1991, \aap, 249, 521 
 
\reference{Hen95} Henning, Th., Begemann, B., Mutschke, H., \& Dorschner, J. 
1995, \aaps, 112, 143 

\reference{Hol94} Holt, S. S., Gotthelf, E. V., Tsunemi, H., \& Negoro, H. 
1994, \pasp, 46, L151
 
\reference{Hou84} Houck, J., R., Shure, M. A., Gull, G. E., \& Herter, T. 1984, 
\apj, 287, L11 
 
\reference{Hur96} Hurford, A. P., \& Fesen, R. A. 1996, \apj, 469, 246 
 
\reference{Ito88} Itoh, H. 1988, \pasj, 33, 1 

\reference{Joh84} Johnston, M. D., \& Yahil, A. 1984, \apj, 285, 587

\reference{Kee93} Keenan, F. P., \& Conlon, E. S. 1993, \apj, 410, 426 

\reference{Kel98} Kelsall, T., et al. 1998, \apj, 508, 44
 
\reference{Keo98} Keohane, J. W., Gotthelf, E. V., \& Petre, R. 1998, \apj, 
503, L175 
 
\reference{Keo96} Keohane, J. W., Rudnick, L., \& Anderson, M. C. 1996, \apj, 
466, 309 
 
\reference{lag96} Lagage, P. O., Claret, A., Ballet, J., Boulanger, F., 
C\'esarsky, C. J., C\'esarsky, D., Fransson, C., \& Pollock, A. 1996, \aap, 
315, L273 

\reference{Mat90} Mathis, J. S. 1990, \araa, 28, 37
 
\reference{Mez86} Mezger, P. G., Tuffs, R. J., Chini, R., Kreysa, E., \& 
Gem\"und, H.-P. 1986, \aap, 167, 145 
 
\reference{Mos93} Moseley, H., Arendt, R., Dwek, E., Casey, S. C., Chan, K. W., 
Glaccum, W. J., Graham, J., Loewenstein, R. F., \& Smith, R. 1993, \baas, 
25, 851 
 
\reference{Pei71} Peimbert, M., \& van den Bergh, S. 1971, \apj, 167, 223 

\reference{Ree95} Reed, J. E., Hester, J. J., Fabian, A. C., \& Winkler, P. F. 
1995, \apj, 440, 706

\reference{Sak92} Saken, J. M., Fesen, R. A., \& Shull, J. M. 1992, \apjs, 
81, 715 
 
\reference{Sea71} Searle, L. 1971, \apj, 168, 41 
 
\reference{Sut95} Sutherland, R. S., \& Dopita, M. A. 1995, \apj, 439, 381 

\reference{Tuf98} Tuffs, R. J., Drury, L. O'C., Rasmussen, I., Heinrichsen, I.,
Innes, D. E., Russell, S., Schnopper, H., \& V\"olk, H. J. 1998, in preparation
 
\reference{vdb71} van den Bergh, S. 1971, \apj, 165, 457 
 
\reference{Vin96} Vink, J., Kaastra, J. S., \& Bleeker, J. A. M. 1996, \aap, 
307, L41
 
\reference{Wea78} Weaver, T. A., Zimmerman, G. B., \& Woosley, S. E. 1978, 
\apj, 225, 1021 
 
\reference{Win91} Winkler, P. F., Roberts, P. F., \& Kirshner, R. P. 1991, in 
Supernovae. The 10th Santa Cruz Workshop in Astronomy and Astrophysics, ed. 
S. E. Woosley, (New York: Springer-Verlag), 652 
 
\reference{Woo84} Woosley, S. E., Axelrod, T. S., \& Weaver, T. A. 1984, in 
Stellar Nucleosynthesis, ed. C. Chiosi \& A. Renzini, 
(Dordrecht: D. Reidel), 263 
 
\reference{Wri80} Wright, E. L., Harper, D. A., Loewnstein, R. F., Keene, J., 
\& Whitcomb, S. E. 1980, \apj, 240, L157 
 
\end{references}
\end{document}